\documentclass[journal]{IEEEtran}

\usepackage[utf8]{inputenc}						% Input encoding
\usepackage{amsmath}							% Math
\usepackage{array}
\usepackage{multirow}
\usepackage{rotating}
\usepackage{url}

\title{Automatic Fado Music Classification} 
\author{Pedro~Girão~Antunes,
        David Martins de Matos,
        Ricardo Ribeiro,
        Isabel Trancoso\\
        Spoken Language Systems Lab (L2F) -- INESC-ID Lisboa, Rua Alves Redol 9, 1000-029 Lisboa, Portugal\\
        \url{pedro.girao.antunes@inesc-id.pt, david.matos@inesc-id.pt, ricardo.ribeiro@inesc-id.pt, isabel.trancoso@inesc-id.pt}}% <-this % stops a space
\begin{document}
\maketitle

\begin{abstract}
In late 2011, Fado was elevated to the oral and intangible heritage of humanity by UNESCO. 
This study aims to develop a tool for automatic detection of Fado music based on the audio signal. 
To do this, frequency spectrum-related characteristics were captured form the audio signal: in addition to the Mel Frequency Cepstral Coefficients (MFCCs) and the energy of the signal, the signal was further analysed in two frequency ranges, providing additional information. Tests were run both in a 10-fold cross-validation setup (97.6\% accuracy), and in a traditional train/test setup (95.8\% accuracy). The good results reflect the fact that Fado is a very distinctive musical style.
\end{abstract}

\section{Introduction}
\begin{figure}[h]
\centering 
\includegraphics[width=.85\columnwidth]{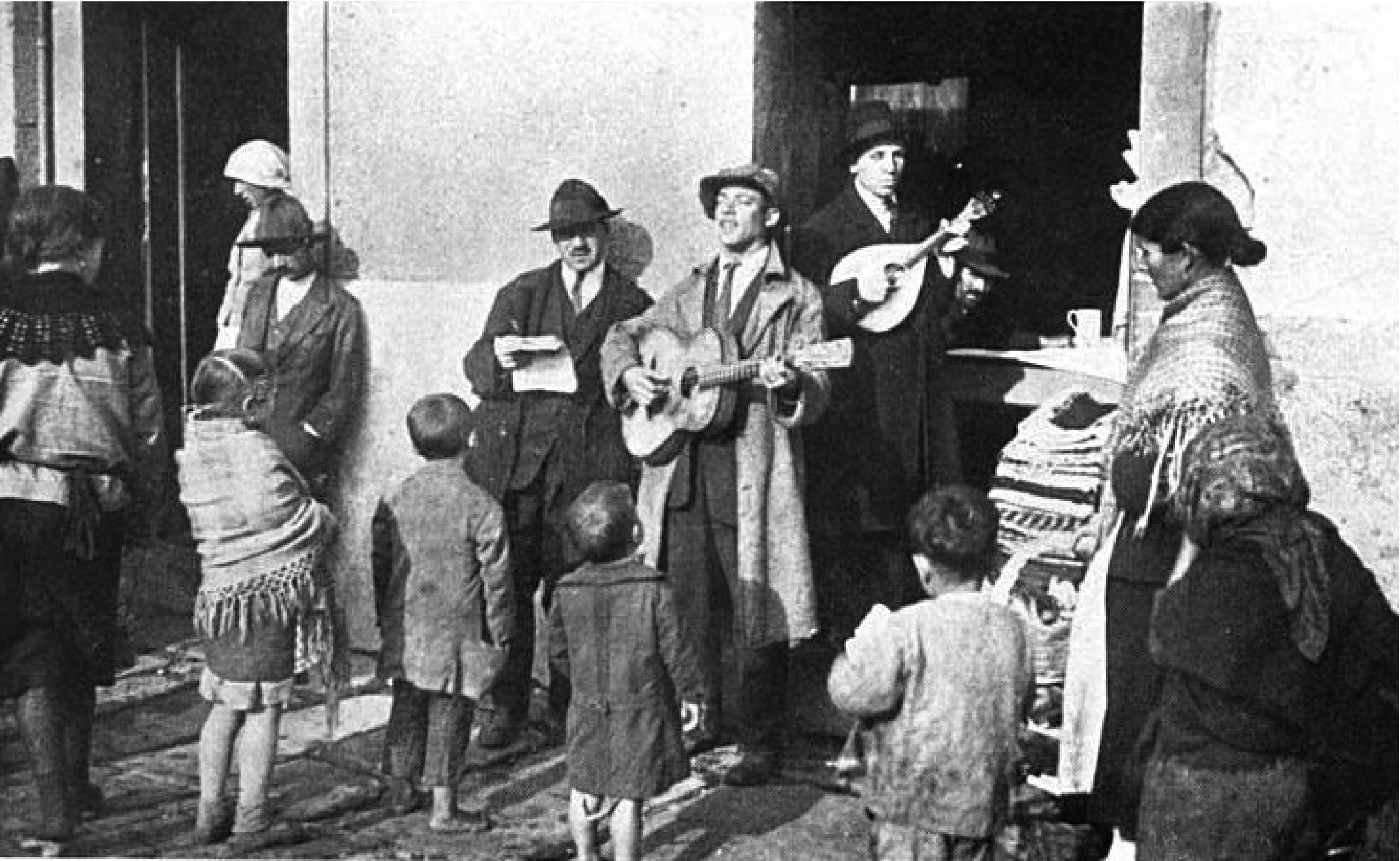} 
\caption{Fado being performed on the streets.}
\label{fado-na-rua}
\end{figure}

Fado music was born in the popular contexts of the 1800s Lisbon, Portugal. It incorporates both music and poetry. 
It was originally introduced among the poorest social groups of the city and it was often performed 
on the streets by amateurs (figure~\ref{fado-na-rua}). It later became an actual profession, and Fado performers -- Fadistas -- now can be found in specialized Casas de Fado (``houses of Fado''), typical Portuguese places dedicated to Fado shows.

Fado songs are typically performed by a solo singer, male or female, traditionally accompanied by a 
classical guitar and the Portuguese guitar -- a pear-shaped cittern with twelve wire strings, unique 
to Portugal. In the past few decades, the instrumental accompaniment was expanded to two Portuguese 
guitars, a classical guitar, and an acoustic bass guitar. The limited instrumental setup helps to better 
define Fado from a signal analysis point of view.
 
This article studies how to automatically determine if a song is Fado or not. It is organized as follows: we start by reviewing relevant related work, both in terms of feature extraction and audio-based music genre classification (section~\ref{relwork}); our Fado classifier is introduced in section~\ref{fadoclass}; the experimental setups, as well as the used dataset, are presented in section~\ref{experiments}; section~\ref{discussion} presents some insights about the classification results; and, finally, section~\ref{conclusions} presents the conclusions.

\section{Related Work}
\label{relwork}

%Music descriptors are absolutely necessary for searching reasons. They include, among others, 
%the artist name, the date, the album name and often the music genre tags. These are called Metadata.

%%%FIXME
The music genre tags are often generated by individual users, this is the cheapest way to do so but can lead to contradictions since people do not have the exact same perception of a given genre. Songs can be manually tagged by experts, however this represents an expensive task, mainly when dealing with big collections of music. Finally, the automatic genre tagging from an acoustic analysis takes the best from the above mentioned methods, as the tagging is done in a consistent way, but it is cheaper to run on large music collections. The problem with this approach is how well a machine can be set to determine the genre of a track. McKay and Fujinaga~\cite{Mckay2006} introduce arguments for and against automatic genre classification. In the next subsection some of the most important works in this area are presented.

\subsection{Automatic Audio-based Music Genre Classification}
Tzanetakis and Cook~\cite{Tzanetakis2002} were among the first to approach the problem of automatic classification of musical genre based on the audio signal. In their work they introduced a dataset called GTZAN also used by Li \cite{Li2003} and Panagakis \cite{Panagakis2009}. Tzanetakis and Cook used features to represent the timbral content, rhythmic content and pitch content and statistical pattern recognition classifier. 

Li et al.~\cite{Li2003} introduced a new feature extraction method for music genre classification. The DWCHs (Daubechies Wavelet Coefficient Histograms) to capture the local and global information of music signals simultaneously by computing histograms on their Daubechies wavelet coefficient. Support Vector Machines and Linear Discriminant Analysis where used to compare the effectiveness of DWCHs with previously used features. It was shown that it improves the accuracy of music genre classification. 

Ahrendt et al.~\cite{Ahrendt2005} used a multivariate autoregressive model of the first 6 MFCCs and a generalized linear model classifier. 

Bergstra et al.~\cite{Bergstra2006} used a variety of timbral related features, including FFT coefficients, MFCCs, zero-crossing rate, among others. Each feature was computed and \textit{m} consecutive frames were grouped in segments. Each segment was then independently classified using AdaBoost. The song is classified according to the ``most voted'' label of its segments.  

Annesi et al.~\cite{Annesi2007}, in addition to the timbral, rhythmic, and pitch features, introduced a new one, which they called Volume Reverse. It consists simply of subtracting the audio wave of a channel to the other one and compute the absolute value. This feature was built upon the claim that each musical genre is recorded differently and that is reflected on the balancing of stereo channels. 

Silla Jr. et al.~\cite{SillaJr.2008} introduced musical genre classification using multiple feature vectors, from the beginning, middle, and end of the song, and a set binary classifiers that were merged to obtain the final genre label. 

Panagakis used Auditory Temporal Modulation and a Sparse Representation-based classifier and obtained the best results on the datasets used so far. However their results were recently contested by Sturm and Noorzad~\cite{Sturm2012}.  

Salamon et al.~\cite{Salamon2010} presented a classifier based on high-level melodic features that are extracted directly from the audio signal of polyphonic music. 

Aryafar and Shokoufandeh~\cite{Aryafar2011} used Explicit Semantic Analysis of textual documents to represent audio samples to feed an support vector machine (SVM) and a k-nearest neighbor (kNN) clustering classifier. 

Table \ref{pw} provides information about the used datasets and features, and the classification results.

\begin{table}[htpb]
\caption{Summary of Automatic Audio-based Music Genre Classification Datasets and Results.}
\label{pw} \centering
\small
\begin{tabular}{p{.07\columnwidth} |  p{.32\columnwidth} | p{.16\columnwidth} | p{.09\columnwidth} | p{.1\columnwidth} }%\hline
 Work     & Dataset & Size & Genres & Results\\\hline%\hline
\cite{Tzanetakis2002}  &GTZAN      &1000   &10   &61\\ 
\hline 
\cite{Li2003}        &GTZAN, \mbox{Local}    &1000, 756 &10, 5   &79, 74\\
\hline
\cite{Ahrendt2005}  &Uspop, \mbox{Magnatune}  &1414, 1515   &6, 10 &78, 61 \\
\hline
\cite{Bergstra2006}      &Uspop, \mbox{Magnatune}  &1414, 1515 &6, 10 &75, 87\\
\hline
\cite{Annesi2007}   &\mbox{Magnatune}, Local    &1515, 500 &10, 5   & 82, 92\\
\hline
\cite{SillaJr.2008}    &Latin Music Database   &3169   &10   &65\\
\hline
\cite{Panagakis2009}   &GTZAN, \mbox{ISMIR 04}   &1000, 1458 &10, 6 & 91, 94\\
\hline
\cite{Salamon2010}    &Local      &500    &5  &90\\
\hline
\cite{Aryafar2011}     &Garageband     & 1886  & 9   & 59\\ 
%\hline
\end{tabular}
\normalsize
\end{table}

\subsection{Evaluation Concerns}

Termens~\cite{Termens2009} notes that the final accuracy varies inversely with the number of genres used, according to the equation:
\begin{equation} \label{genres}
\% \thickspace of \thickspace accuracy = A \times r ^{\#genres}
\end{equation}
In this equation, $A = 93.36$ and $r = 0.9618$, according to the regression method used. The data presented 
in table \ref{pw} is consistent with equation \ref{genres}.
This fact raises the question about the validity of music genre classification, that is, 
how defined in terms of individual perception is genre tagging? For some users, a song is 
considered to be Pop, while for others it can be Rock, Pop-Rock, or any other variety of 
similar tags. This is the cause of the inverse relation between accuracy and number of 
genres considered. Eventually, it would be the same for any manual human tagging~\cite{Scaringella2006}.
Of course, there is also some argument about what is and what is not music. However, 
due to the traditional origin and limited musical instrument in the specific genre, Fado music 
is much more well defined then, for instance, Rock.

%%%%%FIXME
Two works present genre classification in a one against all setup~\cite{Li2003, Aryafar2011}. These works will be useful later, when discussing our results.
 
%In the next subsection the Fado music characteristics, in musical terms and historical is introduced.

\section{Fado Classifier}
\label{fadoclass}

Our Fado classifier was implemented using two main tools: MIRtoolbox~\cite{mirtoolbox} 
for Matlab\r and libsvm~\cite{libsvm}. The first one was used to extract the audio features 
and the second to classify songs. Before introducing the audio features, it must be noted that the audio 
files used were songs in the WAV format, downsampled to 22050Hz, normalized according to the root mean 
square energy (RMS), and that each feature was computed for a 10-second excerpt of each song. The excerpt  
size is discussed in the next section.

\subsection{Audio Features}

We extracted three features: one related to the rhythmic information, one that is related to the timbre, and another related to the musical dynamics.

{\bf The rhythmic feature} was based on the work of Mitri, Uitdenbogerd and Ciesielski~\cite{Mitri2004}. This feature is composed of two 9-dimensional vectors: one referring to low frequencies and another to the high frequencies. The first one is computed based on the FFT coefficients on the 20Hz to 100Hz frequency range, the other is on 8000Hz to 11025Hz. Each set of FFT coefficients was extracted from 50-ms frames, half overlapping. Considering 
this, each component of the 9-dimensional vector is described in table \ref{rf}.

\begin{table}[b]
\caption{Rhythmic feature setup. $v\_in$ is a matrix with the FFT coefficients with frequency 
varying through columns and time varying through lines.}
\label{rf} \centering
\begin{tabular}{l|l}
$maxampv\_in$ 		& max of the average $v\_in$ along time.					\\ \hline
$minampv\_in$ 		& min of the average $v\_in$ along time.					\\ \hline
$count\_80v\_in$	& number of $v\_in$ values above $0.8\thinspace.\thinspace maxampv\_in$.	\\ \hline
$count\_15v\_in$	& number of $v\_in$ values above $0.15\thinspace.\thinspace maxampv\_in$.	\\ \hline
$count\_maxv\_in$	& number of $v\_in$ values above $maxampv\_in$.					\\ \hline
$count\_minv\_in$	& number of $v\_in$ values below $minampv\_in$.					\\ \hline
$meanv\_indist$		& mean distance between peaks.							\\ \hline
$stdv\_indist$		& standard deviation of peaks distance.						\\ \hline
$maxv\_indist$		& max distance between peaks.							\\
\end{tabular}
\end{table}

The purpose of this feature is to capture the rhythmic information present in both, the 
low frequencies and the high frequencies. The fact is that Fado in general 
does not present much information in the low range frequencies, for example, it does not 
have electric bass or drum kick. Moreover, it is certainly richer in terms of high range 
frequencies range, mostly because of the Portuguese guitar. Then, this feature can present 
a good way to distinguish Fado music from other musical genres.

{\bf The timbre feature} is the well known and extensively used MFCCs. We computed the MFCCs based on the FFT coefficients from the whole spectrum. Tests were made in order to use only the mid-range frequencies (from 200Hz to 8000Hz). However, using only this range of frequencies represents a loss of information. The MFCCs were extracted from 50-ms frames, half overlapping, and then the average was computed for each of the 13 components of the vector.

{\bf The dynamic feature} was simply the computation of the RMS energy of the 10-second audio signal before the normalization. 

These three features are then concatenated into a 32-dimensional vector, where the first 
component is the RMS energy, then the two rhythmic features, first the low range one then 
the high range, and finally the MFCCs.

\subsection{Support Vector Machines}

SVMs (Support Vector Machines) are a useful technique for data classification, extensively used in the bibliography~\cite{Li2003, Annesi2007, Aryafar2011}. Basically, SVMs aim at searching for a hyperplane that separates the positive data points and the negative data points with maximum margin. 

SVMs try to map the original training data into a higher (maybe infinite) dimensional space by a function $\phi$. For that, SVMs create a linear separating hyperplane with the maximal margin in this higher dimensional space. From the mathematical point of view, given a training set of instance
label pairs \begin{math} (x_i,y_i),i =1...l \end{math}, where \begin{math} x_i \in R_n \end{math} and \begin{math} y \in {\-1, 1}^l \end{math}, SVMs 
search the solution of the following optimization problem:

\begin{equation} \label{svm1}
\begin{gathered}
min _{w,b,\xi} \thickspace \thickspace \frac{1}{2}w^Tw +C \sum{i=1}^{l}\xi_i \\ 
\text{subject to:} \thickspace  y_i (w^T\phi(x_i)+b)\geq 1-\xi_i; \thickspace \xi>0
\end{gathered}
\end{equation}

Here training vectors $x_i$ are mapped into the higher dimensional space by the function 
$\phi$. $ C> 0 $ is the penalty parameter of the error term. Furthermore, \begin{math} K(x_i,x_j) = \phi(x_i)^T\phi(x_j) \end{math} 
is called the kernel function. There are four basic kernel functions: linear, polynomial, 
radial basis function and sigmoid but, for the music classification problems, the 
Polynomial and Radial Basis Function (RBF) are the most commonly used. 
In our case, the RBF was used as kernel: 

\begin{equation} \label{svm2}
K(x_i , x_j ) = \exp(\-\gamma \parallel x_i \- x_j \parallel ^2 ) \gamma > 0
\end{equation}

In equation~\ref{svm2}, $\gamma$ is a kernel parameter.

The implementation of SVM used was the libsvm, which is an integrated software 
for support vector classification, regression and distribution estimation \cite{libsvm}.

\section{Experiments}
\label{experiments}

In this section, we introduce the experiments made, as well as the dataset used. One of the 
posed questions was ``how long would the sample need to be to characterise the musical genre?''. In 
general, a value from 10 to 30 seconds is used. In practice, we observed that an average person was able 
to recognize the musical genre, in most of the cases, in less then 10 seconds. Thus, we used 10-second music samples.

\subsection{Dataset}

The goal is to classify Fado music specifically. Then, it is about applying a simple binary 
classifier, where one class is Fado and the other is every other musical genre (i.e., not Fado). However, it 
is practically impossible to include every specific single genre and their subgenres in a 
set of songs. Even Fado itself can be divided into several different subgenres~\cite{Gouveia2009}. 
In this case, it makes sense to use as non-Fado, the largest number of different other genres. 
We used ten: Classical, Rock, Jazz, Pop, Folk, Medieval, Blues, Country, Tribal, and Electronic. 
Then, the Fado dataset used is a local set composed by 500 songs: 250 Fado songs and 250 
non-Fado songs.

When we consider 10-second samples, a question poses itself: where should we take it from? We experimented with four hypotheses: from the beginning of the track, from the end, from the middle, or the 10-second segment with the maximum RMS moment.

The first three do not imply extra computation, thus are a simple way of taking the audio samples. However, the fourth one does require some extra computation to determine the maximum RMS moment. In order to do so, the whole song has to be analysed, meaning that the RMS is computed frame 
by frame (50-ms frames, half overlapping). The chosen $10s$ are centered in the maximum RMS frame.

In musical terms, the beginning and end of a song, in general, represent specific and characteristic musical events of the genre in question. For example, a Fado song generally begins with an instrumental introduction, and it ends with an instrumental arrangement. Taking a frame from the middle of the song is an attempt to capture a song in its ``steady state'', that is, with all the instruments and vocals present.

The maximum RMS is an attempt to capture a sample of the song where a relevant event occurs. That would be a refrain entrance, a dynamic change, making use of intensity, and so on. Since Fado tends to be consistent in terms of intensity, the idea is to capture that difference regarding other musical genres.

%(Other genres classification)
%(Address the question of Applying to other genres)

\section{Results and Discussion}
\label{discussion}

The results, using a 10-fold cross validation setup, are shown in table \ref{results1}. The results show that the maximum RMS and the 
beginning of the song are the best samples to distinguish Fado from other musical genres 
using these features. This suggests that Fado indeed has a characteristic beginning and 
that it is relatively more uniform then other musical genres. However, the middle sample
 also gets a high result. 

\begin{table}[h]
\caption{10-fold cross validation for different audio samples}
\label{results1} \centering
\begin{tabular}{l|l}
Beginning   & 96.8 \\ \hline
End       & 86.8 \\ \hline
Middle    & 94.2 \\ \hline
Maximum RMS     & 97.6 \\
\end{tabular}
\end{table}

To analyse song-specific results, the dataset was randomly divided in two 
sets, implementing a traditional train/test setup: approximately 2/3 to train (334 songs) and 1/3 to test (166 songs). The training was done using the parameters $C$ (eq. \ref{svm1}) and $\gamma$ (eq. \ref{svm2}),
estimated using a grid algorithm from libsvm \cite{libsvm}. The results are 
shown in table \ref{results2}. 

\begin{table}[h]
\caption{Traditional train/test setup}
\label{results2} \centering
\begin{tabular}{l|l}
Beginning 	& 95.2 \\ \hline
End 			& 77.1 \\ \hline
Middle 		& 89.8 \\ \hline
Maximum RMS 		& 95.8 \\
\end{tabular}
\end{table}

Analysing the false negatives for the maximum RMS sample, which was the one getting the best 
score, there are 3 misclassified songs (table \ref{confusion}) as non-Fado. From these 3, two can actually 
be considered as non-Fado. They were included among a Fado compilation, yet they can be 
classified as popular Portuguese music for dancing. These two songs include drum beat and 
electric bass, which are not commonly featured in Fado music. The third song is a typical 
Fado song. However, in that particular song, the Portuguese guitar sounds different, almost 
as an electric guitar, what is probably enough to confuse the classifier. 

Among the false positives, there are 4 misclassified songs (table \ref{confusion}). These 4 songs all have 
many things in common. They are medieval music and include two features that make them quite 
similar sounding to Fado: a string instrument that sounds a lot like a Portuguese guitar and 
the vocalist's singing voice is very similar to that of Fado singers.

\begin{table}[h]
\caption{Confusion Matrix.}
\label{confusion} \centering
\begin{tabular}{l|l|c|c|c}
\multicolumn{2}{c}{}&\multicolumn{2}{c}{Actual Class}&\\
\cline{3-4}
\multicolumn{2}{c|}{}&Fado&Non-Fado&\multicolumn{1}{c}{Total}\\
\cline{2-4}
\multirow{2}{*}{Predicted Class}& Fado & $80$ & $3$ & $83$\\
\cline{2-4}
& Non-Fado & $4$ & $79$ & $83$\\
\cline{2-4}
\multicolumn{1}{c}{} & \multicolumn{1}{c}{Total} & \multicolumn{1}{c}{$84$} & \multicolumn{1}{c}{$82$} & \multicolumn{1}{c}{$166$}\\
\end{tabular}
\end{table}

\section{Conclusion}
\label{conclusions}

We introduced an audio-based Fado classifier using an SVM classifier~\cite{libsvm}. We used three sets of features: one to model rhythm, one to model timbre (MFCCs), and one related to music dynamics (RMS). The results are comparable to those presented in~\cite{Li2003, Aryafar2011}, although it is quite difficult to make an accurate comparison, because of the different datasets used, and different features and musical genres in question. However, since the accuracy results are above 90\% (in general) and above 95\% (in our best setups, even considering cross-validation), we think they can be considered state of the art for this kind of task. We introduced comparative study on sample capturing and a detailed analysis on false negatives and false positives. 

This work can be seen as a first step to Fado classification, one possible direction would be to classify different subgenres of Fado music, which would find application in musicological studies. Another possible benefit of this work would be to use this classifier to expand other music classifiers to include Fado classification.

\subsection*{Acknowledgments}

A special thank to Daniel Gouveia, for providing his text on Fado Music and a large collection of Fado audio tracks.

\bibliographystyle{plain}
\bibliography{document}

\end{document}